\documentclass[pra,twocolumn,nofootinbib,floatfix,10pt]{revtex4-2}

\usepackage{amsmath}
\usepackage{amssymb}
\usepackage{wasysym}
\usepackage{graphicx}
\usepackage{color,soul}
\usepackage{physics}
\usepackage{siunitx}
\usepackage{dsfont}
\usepackage{float}
\usepackage{xcolor}
\usepackage[english]{babel}
\usepackage{blindtext}
\usepackage[english,nomargin,inline,marginclue,draft]{fixme}
\pdfpageheight\paperheight
\pdfpagewidth\paperwidth

\usepackage[colorlinks,linkcolor=blue,anchorcolor=blue,citecolor=blue,urlcolor=blue]{hyperref}

\fxusetheme{colorsig}
\FXRegisterAuthor{cg}{acg}{CG}  
\FXRegisterAuthor{th}{ath}{\color{blue}TH}  
\FXRegisterAuthor{ib}{aib}{\color{red}IB} 
\FXRegisterAuthor{sh}{ash}{\color{cyan}SH} 
\FXRegisterAuthor{db}{adb}{\color{green}DB} 
\FXRegisterAuthor{ps}{aps}{PS}
\makeatletter
\renewcommand*\FXLayoutInline[3]{%
  {\@fxuseface{inline}\ignorespaces{\color{fx#1}[#3: #2]}}}
\makeatother

\long\def\symbolfootnote[#1]#2{\begingroup%
\def\thefootnote{\fnsymbol{footnote}}\footnotetext[#1]{#2}\endgroup}

\def\nobreakbefore{%
  \relax\ifvmode\else
    \ifhmode
      \ifdim\lastskip > 0pt\relax
        \unskip\nobreakspace
      \else 
        \nobreakspace
      \fi
    \fi
  \fi
}
\let\oldcite\cite
\renewcommand\cite{\nobreakbefore\oldcite}

\begin{document}
\title{Critical Microwave Mach-Zehnder-Type Interferometry with Dual-LO Rydberg Atoms}

\author{Jun-Rong Chen$^{1,2,4}$}
\author{Guo-Qing Qin$^{3}$}
\author{Peng-Fu Liang$^{3}$}
\author{He Hao$^{3}$}
\author{Ming-Min Zhao$^{2}$}
\author{Ling-Qiang Meng$^{4}$}
\author{Gui-Lan Li$^{3}$}
\author{Min-Jian Zhao$^{2}$}
\author{Bin-Bin Wei$^{5,\textcolor{blue}{\ddagger}}$}
\author{Hao Tian$^{1,\textcolor{blue}{\dagger}}$}

\affiliation{$^1$School of physics, Harbin Institute of Technology, Harbin, Heilongjiang 150001, China.}
\affiliation{$^2$College of Information Science and Electronic Engineering, Zhejiang University, Hangzhou, Zhejiang 310058, China}
\affiliation{$^3$Beijing Institute of Radio Measurement, Beijing 100854, China}
\affiliation{$^4$School of Physics and Opotoelectric Engineering, Hangzhou Institute for Advanced Study, University of Chinese Academy of Sciences, Hangzhou, Zhejiang 310058, China}
\affiliation{$^5$Institute of System Engineering, Tianjin 300161, China.}

\date{\today}

\symbolfootnote[2]{tianhao@hit.edu.cn}
\symbolfootnote[3]{weibb.2009@tsinghua.org.cn}

\maketitle

\textbf{
    High-precision phase measurement of microwave fields underpins a wide range of applications, including wireless communications, distributed radar, plasma diagnostics, and antenna metrology. Existing Rydberg-atom-based approaches, however, often face trade-offs among phase resolution, measurement range, and system complexity. Here we demonstrate a Rydberg-atom-based microwave Mach-Zehnder-type interferometer using a dual-local-oscillator configuration. The two local oscillators establish two coherent interferometric pathways in the Rydberg medium. Their coherent mixing with the signal field produces an interferometric intermediate-frequency output governed by a phase-to-intensity transfer characteristic that enables critical-point enhancement. This scheme supports direct phase retrieval with a resolution exceeding $0.1^\circ$ and unambiguous full $360^\circ$ phase coverage with the reconfigurable dual-LO architecture. Moreover, near the critical interference point, the system exhibits a sharply enhanced phase-to-amplitude transduction, where weak amplitude variations are converted into pronounced phase responses, yielding a sensitivity enhancement exceeding 25 dB. Besides, the same interferometric transfer mechanism enables microwave propagation-distance and polarization metrology, achieving a propagation-distance precision below 20 $\mu$m at 5.7 GHz together with a polarization-angle resolution exceeding $0.1^\circ$. This approach eliminates the need for complex optical configurations and lock-in detection, providing a simple, scalable, and reconfigurable Mach-Zehnder-type quantum microwave interferometry framework for multifunctional high-precision microwave metrology.
}

\section*{Introduction}

    With the rapid advancement of wireless communication and precision measurement technologies, high-fidelity characterization of microwave fields has become a cornerstone for next-generation information systems and fundamental physics research. In particular, high-precision phase measurements of free-space electromagnetic fields are vital in a wide range of applications, including next-generation wireless communications \cite{gong2023holographic, zhao2023reconfigurable, ren2011absolute}, distributed radar systems \cite{janoudi2024distributed, lin2016coherent}, plasma diagnostics \cite{kohn2025electron, cunningham2008use}, and antenna metrology \cite{cutler1947microwave, kim2023phased, sebastian2021explicit, jokinen2024over}. Conventional microwave phase measurement techniques predominantly rely on homodyne and heterodyne interferometry, vector network analyzers, counter-based time-interval methods, etc \cite{abd2025comparative, balestrieri2020review}. The practical utility of these approaches, however, is fundamentally undermined by a persistent trade-off among measurement accuracy, operational bandwidth, and system complexity. This trade-off manifests in the need for costly RF front-ends, elaborate calibration, and frequency-specific antennas. As a result, the measurement flexibility, portability, and spectral coverage of conventional techniques remain severely constrained. A fundamentally different measurement paradigm that combines broadband tunability, high phase precision, and architectural simplicity is therefore highly desirable. 

    Compared with conventional metal-antenna-based microwave field sensors, Rydberg atoms serve as an easily prepared quantum sensing platform offering high sensitivity, broadband spectral coverage, and resilience against electromagnetic interference \cite{schlossberger2024rydberg, allinson2026rydberg, yuan2023quantum, liu2023electric, holloway2022overview, meyer2020assessment}. Leveraging electromagnetically induced transparency (EIT) and Autler-Townes (AT) splitting, a standard heterodyne detection scheme enables microwave field characterization by analyzing the intermediate-frequency (IF) signal generated through mixing between a local oscillator (LO) and the signal field \cite{simons2019rydberg, jing2020atomic, cai2022sensitivity}. The sensitivity of this approach has been enhanced to the $\rm{nV}\cdot\rm{cm}^{-1}\cdot\rm{Hz}^{-1/2}$ level through a range of techniques, including many-body criticality \cite{ding2022enhanced, liu2026enhanced}, cavity enhancement \cite{xiao2024low, liang2025cavity, liu2025cavity}, Doppler-broadening suppression \cite{venu2025three, schlossberger2026fundamental, prajapati2023sensitivity, santamaria2022comparison}, auxiliary optical and microwave fields \cite{qiu2026shortwave, schlossberger2026electromagnetically, liu2022continuous}, and fluorescence readout \cite{prajapati2024investigation, zhang2025advancing}. Within the same heterodyne framework, Rydberg receivers have also demonstrated ultra-broadband operation spanning from near-DC to the THz regime, enabling detection of weak fields across this entire spectral range \cite{yao2026ultra, zhang2024tunable, hu2022continuously, liu2022highly, zhang2024floquet, yuan2024isotropic}. These capabilities have supported measurements of fundamental microwave field parameters \cite{jing2020atomic, simons2019embedding, gordon2019weak, robinson2021angle, elgee2025electrically}, as well as proof-of-concept applications in wireless communications \cite{cui2025rydberg, tong2026broadband, prajapati2022tv, meyer2023simultaneous, nan2026ultra}, radar detection \cite{bohaichuk2022origins, chen2025high}, remote sensing \cite{allinson2026rydberg, arumugam2024remote}, and imaging \cite{watterson2025imaging}. In terms of phase measurement, both all-optical readout schemes based on transient atomic responses and conventional heterodyne detection with a single LO extract the relative phase via time-domain signal fitting, typically achieving precisions around 3 degrees \cite{bohaichuk2026transient, simons2019rydberg, liu2022continuous, jia2021transfer, oliver2026simultaneous}. The incorporation of lock-in detection with optical modulation has further improved phase precision to below 1 degree \cite{jing2020atomic, liu2022all}. To achieve higher phase resolution, closed-loop atomic configurations employing multiple laser fields have been explored, enabling quantum interference-based phase measurements with precisions approaching 0.1 degree \cite{anderson2022optical, han2025phase}. However, such closed-loop interferometric schemes are difficult to generalize to arbitrary signal frequencies and may face limitations in achieving a full 360-degree phase measurement range. Moreover, they significantly increase system complexity, in some cases requiring five or six laser beams at different frequencies, entailing substantial experimental overhead and cost \cite{berweger2023closed}. Consequently, achieving high phase measurement precision while maintaining a simple and scalable system architecture remains a key challenge under existing design frameworks.
    
    In this work, we introduce a Rydberg-atom-based microwave interferometry scheme that realizes a Mach-Zehnder-type interferometer using a dual-local-oscillator (dual-LO) configuration. In this configuration, the two LO fields form effective interferometric arms, while the signal field serves as a common phase perturbation. The Rydberg medium provides a coherent recombination pathway that intrinsically projects the two-arm interference onto the intermediate-frequency (IF) output, yielding a phase-to-intensity transduction function from which the signal phase can be directly retrieved. This approach achieves a phase resolution exceeding $0.1^\circ$, while the reconfigurable dual-LO architecture extends the measurable phase range to the full $360^\circ$. By operating near a critical interference point, the interferometer exhibits a sharply enhanced phase sensitivity, enabling an amplitude-phase-intensity transduction mechanism that yields a sensitivity enhancement exceeding 25 dB. The same interferometric transfer mechanism is further extended to propagation-distance and polarization metrology, achieving a propagation-distance measurement precision below $20~\mu$m and a polarization-angle resolution exceeding $0.1^\circ$ at 5.7 GHz. Our scheme avoids complex optical configurations and modulation-assisted lock-in detection for precision phase, amplitude and polarization angle detection, providing a simple, scalable, and reconfigurable Mach-Zehnder-type microwave interferometry framework for distributed coherent sensing, quantum-enhanced radar, and next-generation wireless communication systems.

    \begin{figure*}
        \centering
        \includegraphics[width=1\linewidth]{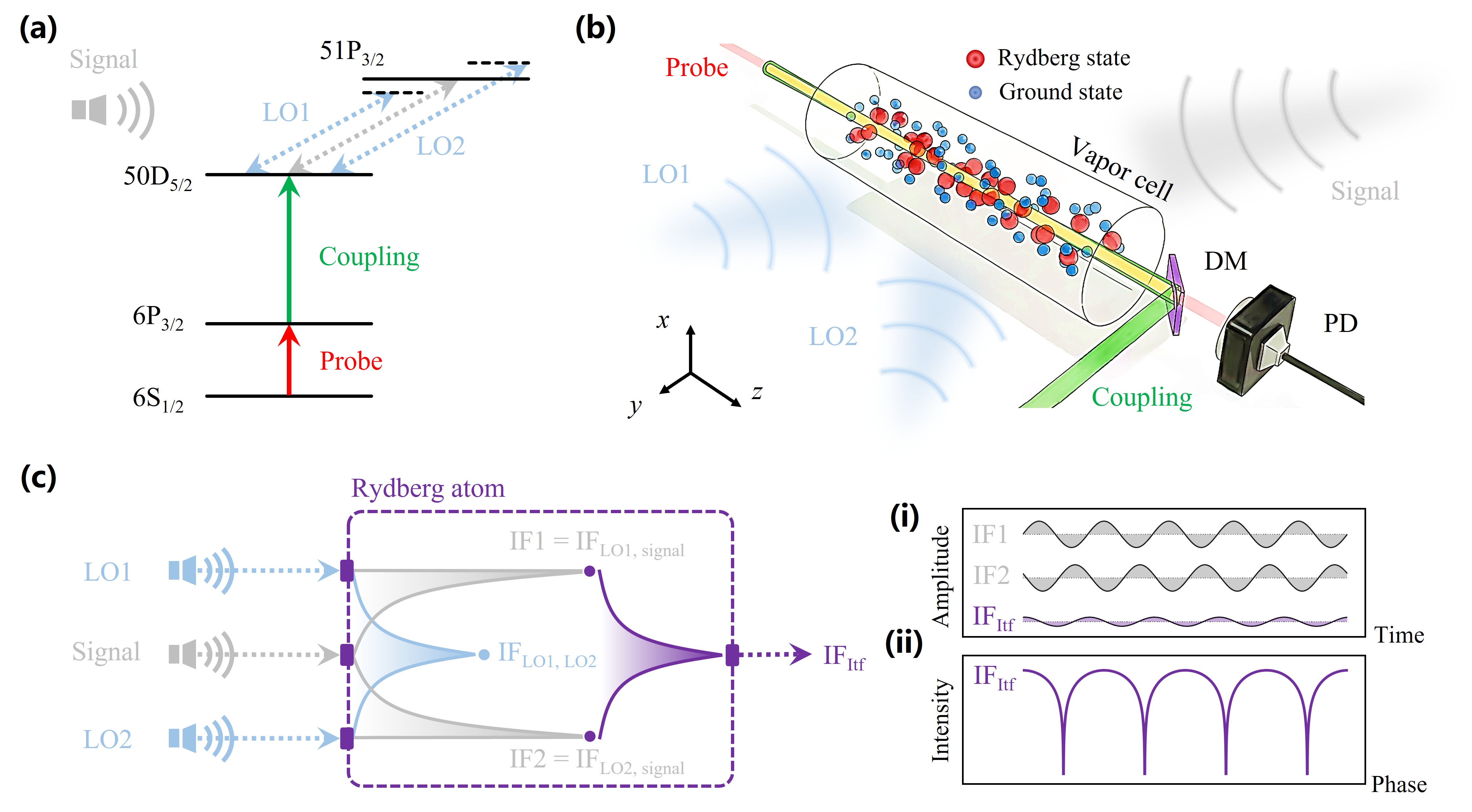}
        \caption{\textbf{Dual-LO Mach-Zehnder-type microwave interferometry.} (a) Energy-level diagram of Cs atoms. A probe beam and a coupling beam drive a two-photon EIT process, exciting Cs atoms from the ground state $6\rm{S}_{1/2}$ to the Rydberg state $50\rm{D}_{5/2}$. Two local oscillator fields (LO1 and LO2) are symmetrically detuned about the signal field frequency and are both near-resonant with the Rydberg transition $50\rm{D}_{5/2} \rightarrow 51\rm{P}_{3/2}$. (b) Schematic of the dual-LO microwave interferometry setup. The counter-propagating probe and coupling beams traverse the atomic vapor cell collinearly. The microwave-induced atomic response is encoded in the transmitted probe field and is read out as the output IF signal by a photodetector (PD). (c) Illustration of the dual-LO microwave interferometry process. The two input local oscillator fields mix with the signal field within the Rydberg ensemble. The resulting degenerate IF signals (IF1 and IF2) coherently interfere within the atomic medium, producing the interferometric output $\rm IF_{\rm Itf}$. Panels (i) and (ii) illustrate the temporal waveform of $\rm IF_{\rm Itf}$ and the corresponding phase-dependent interferometric intensity, respectively.}
        \label{fig1}
    \end{figure*}
    
    \begin{figure*}
        \centering
        \includegraphics[width=1\linewidth]{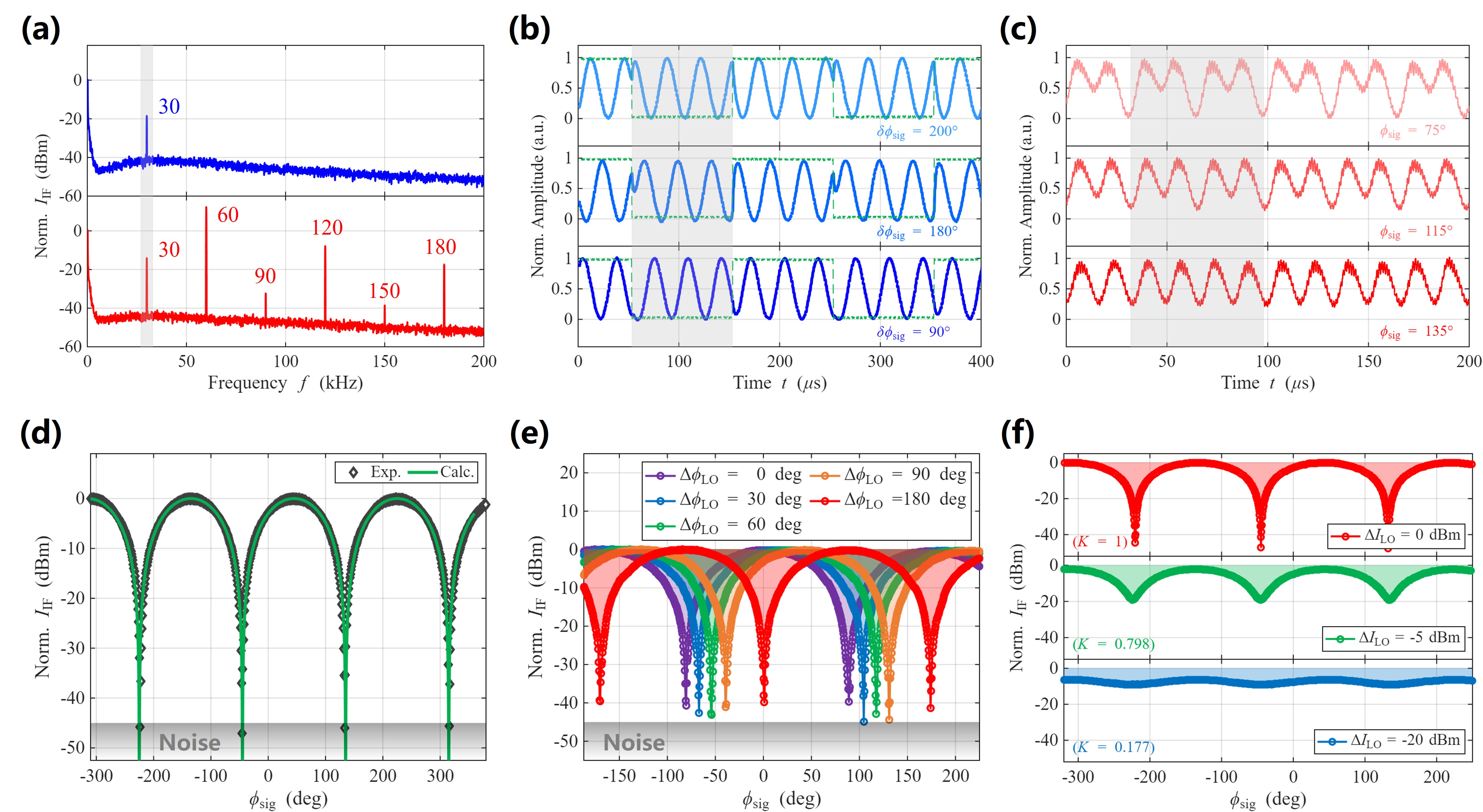}
        \caption{\textbf{Response characteristics of the dual-LO Mach-Zehnder-type microwave interferometer.} (a) Output IF spectra obtained using the conventional single-LO superheterodyne scheme (upper panel) and the proposed dual-LO interferometric scheme (lower panel). (b) Output IF signals under the single-LO standard superheterodyne scheme for varying signal field phase shifts. The upper, middle, and lower panels correspond to phase shifts of 200$^{\circ}$, 180$^{\circ}$, and 90$^{\circ}$, respectively. The green dashed lines indicate the trigger signals for the phase modulation. (c) Output IF signals under the dual-LO interferometry scheme for different signal field phases. The upper, middle, and lower panels correspond to signal phases of 75$^{\circ}$, 115$^{\circ}$, and 135$^{\circ}$, respectively. (d) Response curve of the interference intensity ($I_{\rm{IF}}$) as a function of signal field phase ($\phi_{\rm{sig}}$) in the dual-LO interferometry scheme. Black diamonds denote the experimental data, and the green curve represents the theoretical prediction. (e) Interferometric response curves for different LO phase differences ($\Delta\phi_{\rm LO}$). (f) Interferometric response curves for different LO intensity mismatches ($\Delta I_{\rm LO}$).}
        \label{fig2}
    \end{figure*}

    \begin{figure*}
        \centering
        \includegraphics[width=1\linewidth]{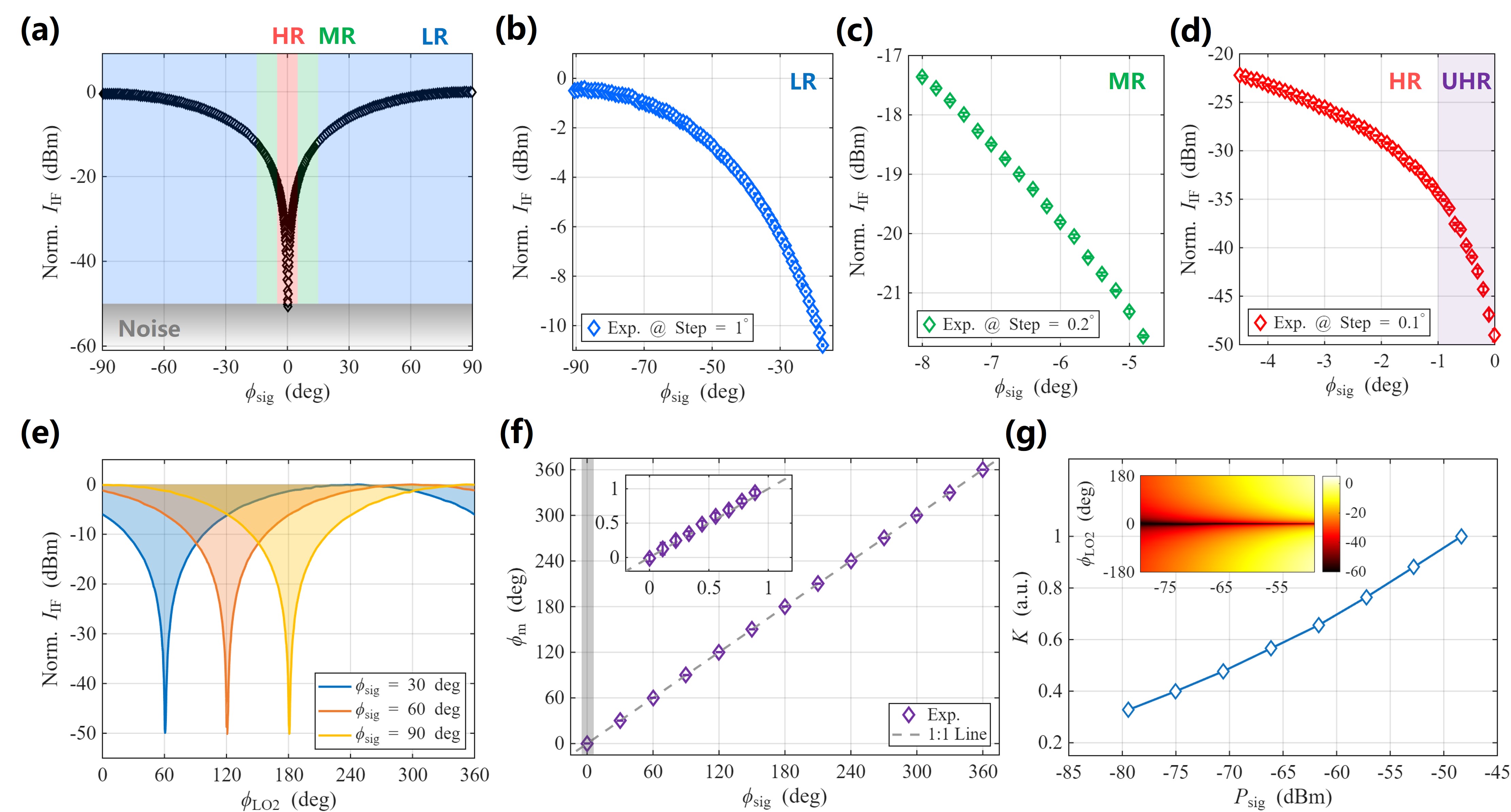}
        \caption{\textbf{High-precision phase retrieval using the dual-LO Mach-Zehnder-type microwave interferometer.}. (a) Interferometric response as a function of signal-field phase over the range from $-90^\circ$ to $+90^\circ$. The phases of LO1 and LO2 are fixed at $0^\circ$ and $180^\circ$, respectively. As the signal phase approaches the critical interference point, the response slope increases significantly. According to the local transfer slope, the response can be divided into low-resolution (LR, blue shaded), medium-resolution (MR, green shaded), and high-resolution (HR, red shaded) regions. (b–d) Enlarged views of the LR, MR, and HR regions, corresponding to phase step sizes of $1^\circ$, $0.2^\circ$, and $0.1^\circ$, respectively. Within the HR region, the response slope continues to increase as the signal phase approaches the critical point, ultimately reaching an ultra-high-resolution (UHR, violet shading) regime with phase resolution exceeding $0.1^\circ$. Error bars represent the standard deviation from five independent measurements. (e) Phase-scanning interferometric readout obtained by sweeping the LO2 phase while fixing the signal relative phase at $30^\circ$ (blue), $60^\circ$ (orange), and $90^\circ$ (yellow), respectively. The minimum of each interferometric readout curve uniquely identifies the signal phase. LO1 is fixed at $180^\circ$. (f) Retrieved signal phase over the full $360^\circ$ phase range. The inset shows retrieval results obtained with a phase step of $0.1^\circ$. The gray dashed line denotes the ideal relation $y=x$. Error bars represent the standard deviation from five independent measurements. (g) Contrast $K$ of the phase-scanning interferometric readout as a function of signal-field intensity. The inset shows the corresponding phase-scanning interferometric readout as a function of both LO2 phase and signal-field intensity.}
        \label{fig3}
    \end{figure*}

    \begin{figure*}
        \centering
        \includegraphics[width=1\linewidth]{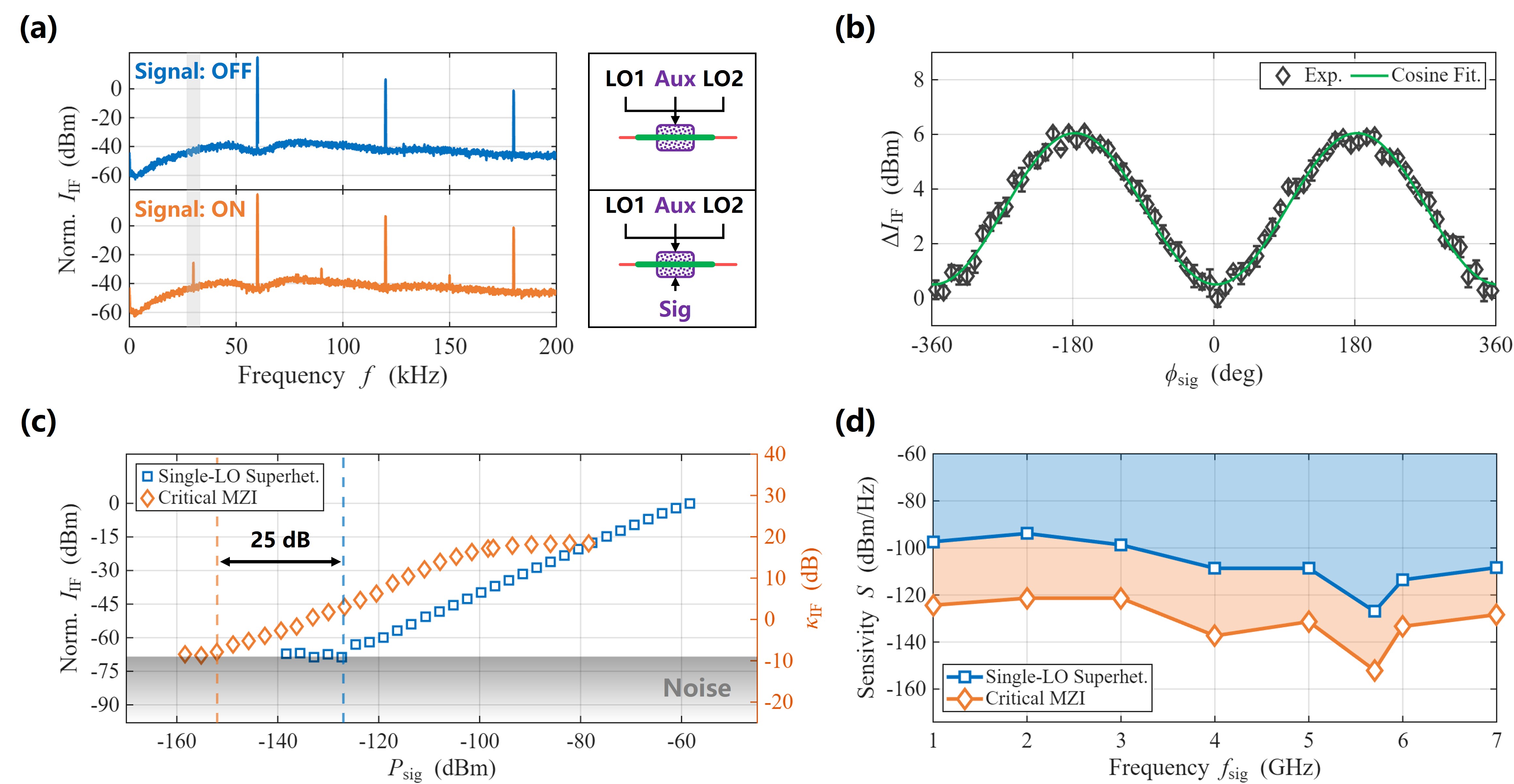}
        \caption{\textbf{Critical-point-enhanced sensitivity in the Mach-Zehnder-type interferometer.}
        (a) Output IF spectra measured with LO1, LO2, and the auxiliary field (Aux) present simultaneously, in the absence (upper panel) and presence (lower panel) of the external signal field. The phases of LO1, LO2, and Aux are fixed at $0^\circ$, $0^\circ$, and $90^\circ$, respectively. The insets schematically illustrate the corresponding operating conditions of the interferometer. The red and green line segments denote the probe and coupling laser fields, respectively, and the purple rectangle represents the atomic vapor cell. Microwave horn antennas are omitted for clarity.
        (b) Relative variation of the interferometric intensity, $\Delta I_{\rm IF}$, as a function of the signal-field phase. The measured response exhibits the expected cosine dependence. Black diamonds represent experimental data and the green curve represents the cosine fit. Error bars indicate the standard deviation from five independent measurements.
        (c) Measured power sensitivity of the conventional single-LO superheterodyne receiver (blue squares) and the proposed critical-point interferometer (orange diamonds) at the resonant operating frequency of 5.7 GHz. The corresponding power sensitivities are $-127$ dBm/Hz and $-152$ dBm/Hz, respectively.
        (d) Measured power sensitivity as a function of operating frequency for the conventional superheterodyne receiver (blue squares) and the critical-point-enhanced Mach-Zehnder interferometer (orange diamonds). The critical-interferometric scheme provides an average sensitivity gain of 24.17 dB over the investigated frequency range, demonstrating that the sensitivity enhancement extends beyond the resonant operating frequency.
        }
        \label{fig4}
    \end{figure*}

    \begin{figure*}
        \centering
        \includegraphics[width=1\linewidth]{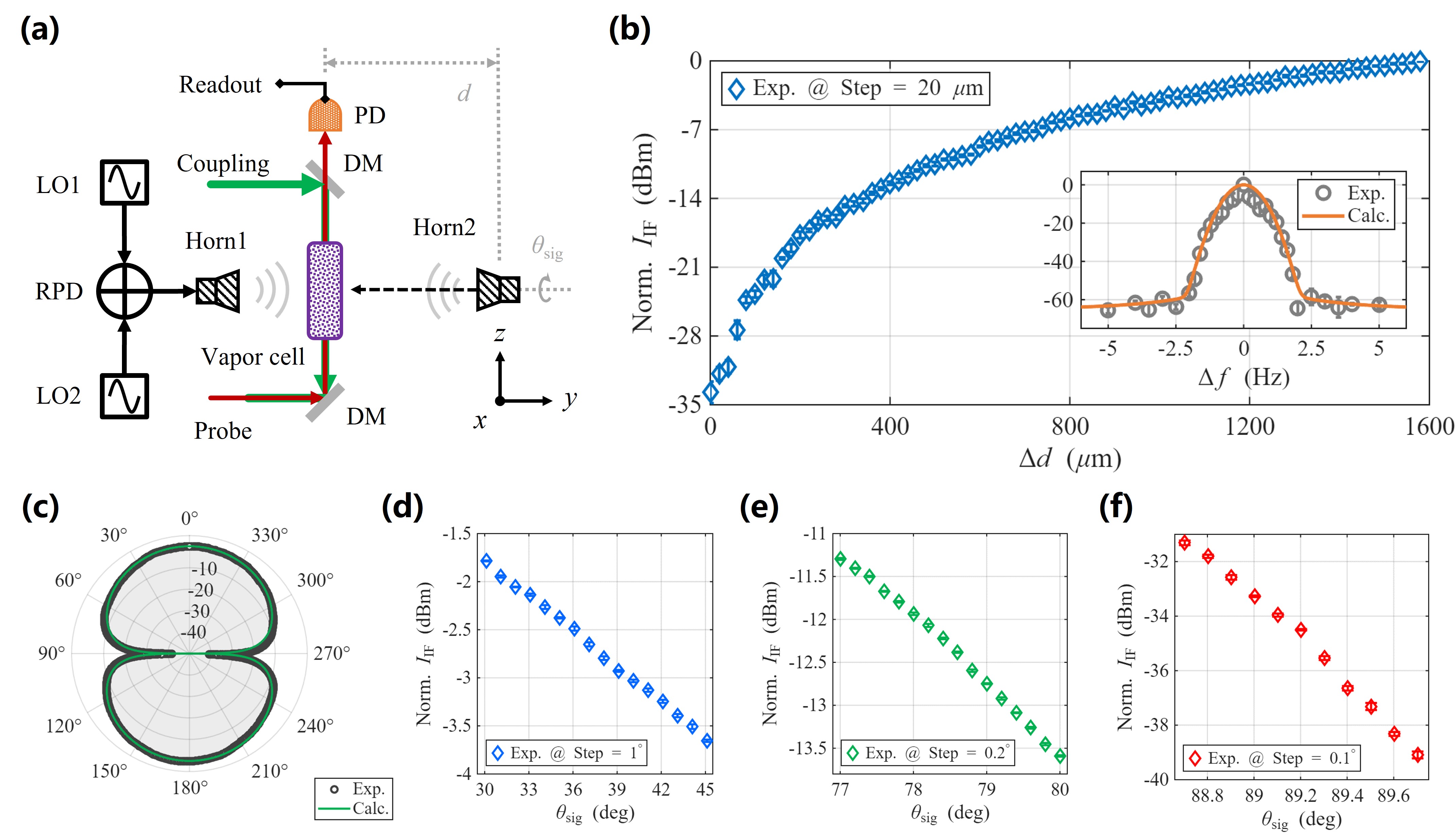}
        \caption{\textbf{Interferometric microwave distance and polarization metrology.}
        (a) Experimental configuration for interferometric microwave metrology. LO1 and LO2 are combined by a resistive power divider (RPD) and transmitted toward the atomic vapor cell through Horn1. The signal field is radiated directly from Horn2 to the vapor cell over a propagation distance $d$. Horn2 can be rotated in the $x$--$z$ plane, defining the signal polarization angle $\theta_{\rm sig}$ with respect to Horn1. The interferometric IF signal is read out from the transmitted probe field by a photodetector (PD).
        (b) Measured interferometric intensity $I_{\rm IF}$ as a function of the distance variation $\Delta d$. Error bars represent the standard deviation from five independent measurements. The inset shows the frequency selectivity of the interferometric response. Gray circles denote the experimental data, while the orange curve is a Gaussian fit. The measured full width at half maximum is approximately 400 mHz.
        (c) Polar representation of the interferometric intensity as a function of the signal polarization angle $\theta_{\rm sig}$. The response exhibits two critical polarization angles within one full rotation. Black circles denote the experimental data, and the green curve represents the theoretical prediction.
        (d-f) Interferometric responses measured at operating points located far from, close to, and in the immediate vicinity of the critical polarization angle, respectively. The polarization angle is varied with step sizes of $1^\circ$, $0.2^\circ$, and $0.1^\circ$ in (d), (e), and (f), respectively. Error bars represent the standard deviation from five independent measurements.}
        \label{fig5}
    \end{figure*}

\section*{Result}

    \textbf{Critical Mach-Zehnder-Type Interferometry Model}
    Conventional heterodyne detection employs a single local oscillator (LO) to down-convert the signal field, encoding its phase and amplitude into an intermediate-frequency (IF) response referenced to the LO. While this approach enables phase retrieval, it lacks intrinsic mechanism for interferometric phase amplification or critical-point operation. Here, we extend this paradigm by introducing a dual-LO configuration that realizes a microwave Mach-Zehnder-type interferometer within a Rydberg atomic medium. Two local oscillator fields, denoted as LO1 and LO2, together with the signal field, simultaneously interact with the atoms, with all fields tuned near the $50\mathrm{D}_{5/2} \rightarrow 51\mathrm{P}_{3/2}$ transition, as illustrated in the energy-level diagram and system schematic of Figs.~\ref{fig1} (a) and (b). The two LO fields are configured with equal amplitudes and symmetric frequency detunings relative to the signal field. In this picture, LO1 and LO2 define two interferometric arms, while the signal field acts as a common phase perturbation applied to both arms. Within the atomic ensemble, multi-wave mixing processes generate two coherent IF components arising from the mixing of the signal field with LO1 and LO2, respectively, as depicted in Fig.~\ref{fig1} (c),
    \begin{equation}
        \mathrm{IF}_1(t) \propto E_{\mathrm{sig}} \cos(\delta\omega t + \phi_{\mathrm{LO1}} - \phi_{\mathrm{sig}}),
    \end{equation}
    \begin{equation}
        \mathrm{IF}_2(t) \propto E_{\mathrm{sig}} \cos(\delta\omega t - \phi_{\mathrm{LO2}} + \phi_{\mathrm{sig}}).
    \end{equation}
    Here, $\phi_{\mathrm{LO1}}$, $\phi_{\mathrm{LO2}}$, and $\phi_{\mathrm{sig}}$ denote the electric-field phases of LO1, LO2, and the signal field, respectively, while $E_{\mathrm{sig}}$ represents the signal amplitude. The angular frequency offset is defined as $\delta\omega = |\omega_{\mathrm{LO1}} - \omega_{\mathrm{sig}}| = |\omega_{\mathrm{LO2}} - \omega_{\mathrm{sig}}|$, where $\omega$ denotes the angular frequency. 
    
    These two components constitute the field amplitudes of the two interferometric paths and are coherently recombined within the same atomic medium. The resulting total IF field is given by the superposition $\rm{IF}_{\rm{Itf}} = \rm{IF}_1 + \rm{IF}_2$, and the detected signal corresponds to its intensity. Assuming equal amplitudes for the two LO fields, the interference term yields
    \begin{equation}
        \begin{aligned} 
            \rm{IF}_{\rm{Itf}} & \propto E^{2}_{\rm{sig}} \left[ 1 + \cos \left( \phi_{\rm{Itf}} \right) \right] 
        \end{aligned} 
        \label{eq3}
    \end{equation}
    where $\phi_{\rm{Itf}} = \phi_{\rm{LO1}} + \phi_{\rm{LO2}} - 2\phi_{\rm{sig}}$ represents the total phase accumulated in the dual-LO interference process. This expression explicitly reveals the Mach-Zehnder-type interference, where the relative phase accumulated along the two effective paths determines the output intensity.
    
    Destructive interference occurs when the phase condition
    \begin{equation}
        \phi_{\mathrm{sig}}^{(c)} = \frac{1}{2}(\phi_{\mathrm{LO1}} + \phi_{\mathrm{LO2}} + \pi + 2k\pi), \quad k \in \mathbb{Z}
        \label{eq4}
    \end{equation}
    is satisfied, defining a critical point of the interferometer. In the vicinity of this point, the output intensity exhibits a steep dependence on the signal phase, enabling efficient transduction of small phase variations into large amplitude changes.

    To stabilize the operation at this critical point, a weak auxiliary field (Aux, omitted from Fig.~\ref{fig1} for clarity), frequency-degenerate with the signal field, is introduced as a bias field. By setting its phase to $\phi_{\mathrm{Aux}} = \phi_{\mathrm{sig}}^{(c)}$, the interferometer is biased at the condition of near-complete destructive interference. A weak incoming signal then perturbs this balance, inducing a small phase shift that lifts the destructive interference and produces a measurable output (see Methods). Denoting the amplitude ratio as $x = E_{\mathrm{sig}} / E_{\mathrm{Aux}}$ with $x < 1$, the maximum induced phase deviation $\delta\phi_{\mathrm{Aux}}$ satisfies
    \begin{equation}
        \delta\phi_{\mathrm{Aux}} = \arctan{\left[ \frac{\tan(\phi_{\rm{Aux}}) + x}
                      { 1 - x \tan(\phi_{\rm{Aux}})} \right]}.
    \end{equation}
    This cascade-amplitude-to-phase conversion followed by phase-to-intensity readout enables substantial enhancement of field detection sensitivity.

    Finally, the phase sensitivity near the critical point can be directly mapped to propagation-distance measurement precision. A minimum resolvable phase shift $\delta\phi_{\mathrm{min}}$ corresponds to a displacement $\delta d_{\mathrm{min}}$ via
    \begin{equation}
        \delta\phi_{\mathrm{min}} = \frac{2\pi}{\lambda} \delta d_{\mathrm{min}},
        \label{eq6}
    \end{equation}
    establishing a direct link between interferometric phase detection and high-precision distance measurement.

    \textbf{Interferometric Transfer Characteristics}
    We first compare the transfer characteristics of the proposed Mach-Zehnder-type interferometer with those of a conventional single-LO superheterodyne receiver. Figure~\ref{fig2} (a) contrasts the representative IF spectra obtained using the two schemes, both measured with a frequency offset of 30 kHz between the signal and LO fields. In the conventional receiver, the signal field mixes with a single LO to produce one IF component. Although this IF signal preserves the relative phase between the signal and the LO, the phase information remains encoded only in the temporal waveform. Consequently, determination of the relative phase requires an additional phase reference, such as the LO clock or an external trigger synchronized with the measurement. Without such a reference, the recorded IF waveform only reveals temporal waveform shifts, making quantitative determination of the relative phase difficult. This behavior is illustrated by the representative IF spectrum and time-domain traces in Figs.~\ref{fig2} (a) and (b), where phase changes are reflected only through waveform shifts and become increasingly difficult to resolve for small phase perturbations.

    In contrast, introducing a second LO fundamentally transforms the phase readout mechanism. The two symmetrically detuned LO fields generate a pair of coherent IF components that interfere within the Rydberg medium, thereby realizing an effective Mach-Zehnder-type interferometer. Besides the higher-order spectral components arising from nonlinear atomic mixing, the interferometric IF component at $f_{\rm IF}$ is formed through coherent recombination of the two interferometric arms and directly encodes the relative phase of the signal field into its intensity. As a result, the interferometer establishes a phase-to-intensity transduction mechanism, allowing the relative signal phase to be determined solely from the recorded IF intensity without requiring any external phase reference. This behavior is evident in Fig.~\ref{fig2} (c), where varying the signal phase produces pronounced variations in the interferometric IF intensity (gray shading) while the temporal waveform itself no longer serves as the measurement observable.

    The resulting interferometric transfer characteristic is presented in Fig.~\ref{fig2} (d). The IF intensity exhibits a periodic dependence on the signal phase, with complete destructive interference occurring at multiple critical points separated by $180^\circ$. Near these critical points, the transfer curve becomes exceptionally steep, enabling minute phase variations to produce large changes in the measured IF intensity. This transfer characteristic forms the physical basis for the high-precision phase retrieval demonstrated in the following section. Similar to a conventional Mach-Zehnder interferometer, the critical point can be continuously tuned by adjusting the relative phase of the two local oscillators, which effectively changes the phase difference between the two interferometric arms. In the present implementation, this is accomplished electronically through the phase difference between LO1 and LO2. According to Eq.~\ref{eq3}, varying the inter-LO phase difference shifts the entire transfer curve along the signal phase axis without altering its shape, allowing the critical interference point to be positioned at an arbitrary signal phase, as verified experimentally in Fig.~\ref{fig2} (e).

    Furthermore, as in a conventional Mach-Zehnder interferometer, optimal interference requires balanced amplitudes in the two interferometric arms. We quantify the interference quality using the fringe contrast $K = (I_{\rm{IF}}^{\rm{max}} - I_{\rm{IF}}^{\rm{min}}) / (I_{\rm{IF}}^{\rm{max}} + I_{\rm{IF}}^{\rm{min}})$. Figure~\ref{fig2} (f) shows that increasing the LO intensity imbalance between the two LOs progressively reduces the fringe contrast and eventually eliminates the critical point. This behavior confirms the interferometric origin of the observed transfer characteristic and experimentally validates the proposed dual-LO interferometric measurement scheme.

    \textbf{High-Precision Interferometric Phase Retrieval}
    The interferometric transfer characteristic established above naturally provides a route toward high-precision phase retrieval. As shown in Fig.~\ref{fig3} (a), the response is periodic with a period of $180^\circ$, with each period containing a critical interference point at which the transfer slope reaches its maximum. Since the local transfer slope determines the phase-to-intensity conversion efficiency, different operating regions exhibit markedly different phase sensitivities. According to the slope magnitude, the response can be approximately divided into three operating regimes, referred to as the low-resolution (LR), medium-resolution (MR), and high-resolution (HR) regions. Figures~\ref{fig3} (b)-(d) experimentally verify the phase discrimination capability in these three operating regimes. Phase increments of $1^\circ$, $0.2^\circ$, and $0.1^\circ$ can be clearly resolved in the LR, MR, and HR regions, respectively, confirming the progressively enhanced phase sensitivity predicted by the transfer characteristic. As the operating point approaches the critical interference point, the transfer slope continues to increase, giving rise to an ultra-high-resolution (UHR) regime, highlighted by the violet shading in Fig.~\ref{fig3} (d), in which the achievable phase resolution exceeds $0.1^\circ$.

    To adopt this transfer characteristic for practical phase measurement, a readout strategy is required. It uniquely maps the unknown signal phase onto the interferometric response. To this end, we implement a phase-scanning interferometric readout by sweeping the phase of LO2 while keeping LO1 fixed. As shown in Fig.~\ref{fig3} (e), this procedure generates a family of interferometric readout curves whose minima, referred to as the critical points, correspond to the critical interference condition. At the critical point, the interference condition is uniquely determined by the relative phase between the signal field and the two local oscillators, thereby establishing a well-defined correspondence between the LO2 phase at the intensity minimum and the signal-field phase within the operating branch. This readout requires no external phase reference or synchronized trigger, since all phase information is intrinsically encoded in the interferometric intensity. However, owing to the $2\phi_{\rm sig}$ dependence in Eq.~\ref{eq3}, the transfer curve exhibits an intrinsic $180^\circ$ ambiguity. Nevertheless, this ambiguity can be removed straightforwardly by exploiting the reconfigurability of the dual-LO architecture. By operating the system in the single-LO configuration, the IF phase resolves the branch ambiguity of the interferometric response. This coarse determination is subsequently refined by the phase-scanning interferometric readout, enabling unambiguous phase retrieval over the full $360^\circ$ range. The performance of the proposed readout strategy is experimentally validated in Fig.~\ref{fig3} (f). Over the entire $0^\circ$-$360^\circ$ range, the retrieved phase agrees closely with the applied phase and follows the ideal $y=x$ relation. The inset further demonstrates measurements performed with a phase increment of $0.1^\circ$, where adjacent phase states remain clearly distinguishable. Error bars correspond to the standard deviation of five independent measurements, confirming both the accuracy and reproducibility of the proposed phase retrieval method.

    The applicability of the proposed method to weak signal fields is evaluated in Fig.~\ref{fig3} (g). As the signal power decreases, the fringe contrast of the phase-scanning interferometric readout curves is gradually reduced, as illustrated by both the measured contrast and the corresponding color map. Nevertheless, the critical interference point remains well defined over the entire investigated power range. Since phase retrieval depends only on locating this critical point rather than on the absolute interference intensity, reliable phase estimation is maintained even for weak signal fields. These results demonstrate that the proposed readout strategy maintains high phase precision over a broad dynamic range of the signal strength, highlighting its suitability for practical weak-field microwave phase sensing.

    \textbf{Critical-Point Sensitivity Enhancement}
    The high phase sensitivity established at the critical interference point naturally extends beyond precision phase retrieval. As we show below, the same critical interference point can also be exploited to substantially enhance the sensitivity of weak microwave-field detection. By operating the interferometer at this critical point, an extremely small perturbation of the interference condition produces a disproportionately large variation in the interferometric intensity, allowing weak microwave fields to be detected through an amplitude-phase-intensity transduction process. To establish this operating condition, we introduce a weak auxiliary microwave field (Aux) at the same frequency as the signal field. The auxiliary field prepares the interferometer to the critical point, where the interferometric component $\rm IF_{\rm Itf}$ is driven below the measurement noise floor, as shown in the upper panel of Fig.~\ref{fig4} (a). When a weak signal field is subsequently introduced, its coherent superposition with the auxiliary field perturbs the critical interference condition and immediately restores a measurable interferometric response, as illustrated in the lower panel. This amplitude-phase-intensity transduction directly converts the critical phase sensitivity of the interferometer into enhanced sensitivity to weak microwave fields.

    To verify this transduction experimentally, we first measure the interferometric intensity variation $\Delta I_{\rm IF}$ by sweeping the  signal-field phase whose power is 50 dB lower than that of the auxiliary field. As shown in Fig.~\ref{fig4} (b), the measured response follows the expected cosine dependence (see Methods), confirming the proposed amplitude-phase-intensity transduction mechanism. In practical measurements, however, the phase of the unknown signal field is not swept directly. Instead, an equivalent interferometric response is obtained by synchronously sweeping the auxiliary-field phase and the phase of LO2 while maintaining the critical phase relation described by Eq.~\ref{eq4}. The resulting readout yields an interferometric response equivalent to that obtained by directly sweeping the signal phase and is therefore adopted throughout the following sensitivity measurements.

    Figure~\ref{fig4} (c) compares the measured power sensitivity of the proposed critical-interferometric scheme with that of a conventional single-LO superheterodyne receiver under identical experimental conditions. For the proposed interferometric scheme, the weak-field response is quantified by the oscillation factor $\kappa_{\rm IF}=10\log(K)$, whereas the conventional receiver is evaluated directly from the IF signal intensity. Operating at the critical point shifts the detectable power limit from $-127$ dBm to $-152$ dBm at 1 Hz frequency resolution bandwidth, corresponding to an improvement of approximately 25 dB in power sensitivity. These results demonstrate that critical-point operation enables reliable detection of substantially weaker microwave fields. Furthermore, we note that the sensitivity enhancement demonstrated above is not restricted to the resonant frequency at which the interferometer is biased. Figure~\ref{fig4} (d) compares the measured power sensitivities of the two detection schemes from 1 to 7 GHz. Across the entire investigated frequency range, the critical-interferometric scheme consistently provides an average improvement of 24.2 dB over the conventional single-LO receiver. The accessible frequency span in the present experiment is limited primarily by the available microwave sources rather than by the interferometric detection principle. These results demonstrate that critical-point interferometry provides a general route for converting interferometric phase sensitivity into substantially enhanced microwave power sensitivity, establishing a practical framework for ultrasensitive broadband Rydberg microwave sensing.
    
    \textbf{Interferometric Microwave Distance and Polarization Metrology}
    The demonstrated interferometric phase sensitivity further enables high-precision microwave distance metrology through a distance-phase-intensity transduction mechanism. As in conventional radar systems, the propagation distance of a microwave field is encoded in its accumulated phase, which is subsequently converted into interferometric intensity through the critical interferometric response. To verify this capability, we implement the experimental configuration illustrated in Fig.~\ref{fig5} (a), where the signal field propagates directly from Horn2 to the atomic vapor cell over a distance $d$. By continuously varying the propagation distance while recording the interferometric intensity extracted from the IF signal, we obtain the experimental results shown in Fig.~\ref{fig5} (b). The measured interferometric response clearly resolves successive distance variations throughout the investigated range, corresponding to a distance measurement precision of approximately $20~\mu$m for a 5.7 GHz microwave field. These measurements experimentally verify the proposed distance-phase-intensity transduction mechanism, whereby propagation-induced phase accumulation is directly mapped onto the interferometric intensity to enable high-precision microwave distance metrology. The measured distance resolution agrees well with the value inferred from the experimentally demonstrated phase resolution of $0.1^\circ$ using Eq.~\ref{eq6}, confirming the consistency between phase and distance metrology in the proposed interferometric framework. Such propagation-distance resolution provides a promising capability for high-precision microwave ranging and sub-wavelength path-length variation measurement.

    Beyond the demonstrated distance measurement precision, the proposed interferometric scheme possesses an intrinsic frequency selectivity that further benefits practical sensing applications. As shown in the inset of Fig.~\ref{fig5} (b), the measured interferometric response has a full width at half maximum (FWHM) of approximately 400 mHz, indicating that frequency offsets exceeding approximately $\pm 200$ mHz rapidly degrade the interference visibility. However, this frequency selectivity is subject to two well-defined degenerate frequency channels. Microwave fields occupying these two channels generate the same IF component as the desired signal and can therefore perturb the interferometric readout. Apart from these two degenerate frequency channels, all other microwave fields fail to satisfy the interferometric condition and are therefore intrinsically rejected by the interferometric readout. Consequently, the interferometer inherently provides a narrowband filtering capability, substantially improving its robustness against out-of-band electromagnetic interference. 

    We further investigate the polarization response of the proposed interferometer. As shown in Fig.~\ref{fig5}(c), the interferometric intensity exhibits a periodic dependence on the signal polarization angle, following a transfer characteristic analogous to that of the phase response, with two critical polarization angles appearing within one full rotation. The slight deviations from the ideal response are mainly attributed to the unavoidable propagation-phase variation introduced by a small height mismatch between Horn1 and Horn2 during antenna rotation. Consequently, progressively enhanced polarization sensitivity is achieved as the operating point approaches the critical polarization angle. The experimental results in Fig.~\ref{fig5} (d)-(f) verify that polarization-angle increments of $1^\circ$, $0.2^\circ$, and below $0.1^\circ$ can be resolved in different operating regions, demonstrating critical-point-enhanced polarization metrology with sub-$0.1^\circ$ angular resolution. Such angular resolution provides a promising capability for precision antenna alignment and microwave polarization calibration. We note that this polarization transfer characteristic suggests an even higher achievable angular resolution in the immediate vicinity of the critical point (see Methods). In practice, however, the attainable resolution is ultimately constrained by the finite precision of the experimental motion control.

    Together with the preceding demonstrations of high-precision phase retrieval and ultrasensitive weak-field detection, these results establish the proposed Rydberg interferometer as a comprehensive microwave metrology platform, in which the phase, amplitude, propagation distance, and polarization angle of the microwave signal are all transduced into a common interferometric intensity observable.

\section*{Discussions and Summary}
    The results presented above reveal a fundamental distinction between the proposed interferometric receiver and conventional single-LO superheterodyne detection. In conventional superheterodyne receivers, the amplitude, frequency, and phase of the microwave field are extracted through waveform analysis of the intermediate-frequency signal. In contrast, the proposed interferometer introduces an interferometric transfer mechanism, whereby microwave-field information is retrieved through interferometric intensity rather than waveform evolution. This interferometric transduction provides an additional degree of freedom for information retrieval and establishes a unified framework for converting different microwave observables into a common interferometric intensity readout. More generally, the proposed interferometric transduction framework is not restricted to any specific microwave observable. As demonstrated in this work, it supports phase retrieval through phase-to-intensity transduction, weak-field sensing through amplitude-phase-intensity transduction, distance metrology through distance-phase-intensity transduction, and polarization metrology through polarization-phase-intensity transduction.
    
    The enhanced performance of the proposed interferometer originates from operation in the vicinity of the critical interference point, where the transfer function exhibits its maximum local slope. Therefore, the attainable phase precision improves continuously as the operating point of the receiver approaches the critical interference point, exceeding the limit achievable with conventional linear phase readout. However, this improvement is accompanied by a reduced high-sensitivity operating range, since the enhanced phase sensitivity is confined to the immediate vicinity of the critical point. To overcome this limitation, we employ the phase-scanning readout strategy introduced in Fig.~\ref{fig3} (e), in which the signal phase is determined from the location of the critical point rather than from a single operating point on the transfer curve. In practice, a complete phase scan is unnecessary. Owing to the symmetry of the interferometric transfer characteristic, the critical-point position can be accurately determined from a limited number of measurements acquired in its vicinity, preserving the high measurement precision while substantially improving the practicality of the readout.

    A similar trade-off exists in the critical-point sensitivity enhancement scheme. By biasing the interferometer at the critical interference condition, substantially weaker microwave fields become detectable than in conventional Rydberg superheterodyne detection. This sensitivity enhancement is achieved at the expense of a reduced dynamic range of approximately 30 dB. For sufficiently strong microwave fields, the induced perturbation drives the interferometer away from the critical operating region, where the transfer characteristic becomes progressively less sensitive. Consequently, the interferometric modulation gradually deviates from the ideal sinusoidal response, leading to saturation of the oscillation factor $\kappa_{\rm IF}$. This trade-off between sensitivity and dynamic range is an inherent consequence of operating near the critical point and is common to many interferometric sensing systems. Depending on the application, the operating point can therefore be flexibly selected to balance sensitivity and dynamic range.

    In summary, we have introduced and experimentally demonstrated a Rydberg-atom Mach-Zehnder-type microwave interferometer based on a dual-LO architecture. Rather than relying on conventional waveform analysis of the intermediate-frequency signal, the proposed approach establishes an interferometric transduction framework in which microwave-field information is retrieved through the interferometric intensity. More generally, the proposed interferometric transduction framework is not restricted to any specific microwave observable. Within this unified framework, we demonstrate high-precision interferometric phase retrieval with a phase resolution exceeding $0.1^\circ$ over the full $360^\circ$ range, critical-point sensitivity enhancement yielding an approximately 25 dB improvement in microwave power sensitivity, interferometric distance metrology with a measurement precision of approximately $20~\mu$m, and polarization metrology with a polarization-angle resolution exceeding $0.1^\circ$. The proposed architecture requires neither complex optical interferometers nor modulation-assisted lock-in detection, while remaining compact, reconfigurable, and readily compatible with existing Rydberg superheterodyne receivers. These results establish interferometric transduction as a versatile new operating principle for quantum microwave sensing and provide a comprehensive platform for multifunctional microwave metrology, encompassing precision phase retrieval, propagation-distance measurement, polarization metrology as well as ultrasensitive weak-field detection, , with promising applications in distributed coherent sensing, quantum radar, and next-generation wireless communication systems.

\section*{Methods}
    
    \textbf{Experimental Setup}
    In our experiments, cesium ($^{133}$Cs) atoms contained in a room-temperature vapor cell are excited to a Rydberg state via a two-photon EIT scheme. This excitation is realized using a probe laser at 852 nm (Toptica DL Pro 852) and a coupling laser at 509 nm (Toptica TA-SHG Pro 510), addressing the $6\mathrm{S}_{1/2} \rightarrow 6\mathrm{P}_{3/2} \rightarrow 50\mathrm{D}_{5/2}$ ladder-type transition. The corresponding Rabi frequencies are $\Omega_p = 2\pi \times 6.87$ MHz and $\Omega_c = 2\pi \times 2.29$ MHz for the probe and coupling fields, respectively, while the probe transmission is detected using a balanced photodetector (Thorlabs PDB210A/M). Microwave fields are supplied by an analog signal generator (Keysight AP5032A) serving as the local oscillator (LO) and a vector signal generator (Rohde \& Schwarz SMW200A) producing the test signal. Both fields are radiated into the vapor cell via dual-ridged horn antennas (HENGDA HD-10200DRHA10S), where they interact with the Rydberg atoms. The resulting atomic response is imprinted onto the probe transmission and subsequently analyzed in the frequency domain, with the IF spectra recorded using a spectrum analyzer (Rohde \& Schwarz FSV3004). For distance measurement, the position of the transmitting antenna is precisely controlled by a motorized linear translation stage (Thorlabs LTS300), which provides a travel range of 300 mm and a minimum incremental motion of 0.1~$\mu$m, enabling calibrated variation of the microwave propagation distance. For polarization measurement, the transmitting horn antenna is mounted on a motorized rotary stage (Zaber XRSW60A) with an angular accuracy of $0.08^\circ$, enabling precise control of the signal polarization angle. Owing to the approximately linear polarization of the microwave field radiated by the dual-ridged horn antenna, rotating the antenna directly changes the polarization angle of the incident signal field.

    \textbf{Dual-LO Interferometric Detection}
    To quantify the phase sensitivity of the proposed interferometer, we first derive its interferometric transfer function. Using the interference model derived in Eq.~\ref{eq3}, the normalized intensity of the interferometric component can be written as
    \begin{align}
        I_{\rm{Itf}} = 10\log \left[ 1 + \cos (\phi_{\rm{Itf}} ) \right], 
        \label{eq7}
    \end{align}
    where $\phi_{\rm{Itf}} = \phi_{\rm{LO1}} + \phi_{\rm{LO2}} - 2\phi_{\rm{sig}}$ defines the effective interferometric phase determined by the two local oscillators and the signal field. Differentiating Eq.~\ref{eq7} with respect to the signal phase gives the local transfer slope
    \begin{align}
        k_{\rm{IF}} = \frac{\rm{d} \it{I}_{\rm{Itf}}}{\rm{d} \it{\phi_{\rm{sig}}}}  = \frac{20}{\ln 10}\tan\left( \frac{\phi_{\rm{Itf}}}{2} \right),
        \label{eq8}
    \end{align}
    The transfer slope increases rapidly as the critical point approaches the critical interference condition, providing the theoretical origin of the experimentally observed enhancement in phase sensitivity. Although Eq.~\ref{eq8} predicts a divergent slope at the critical point, the attainable phase sensitivity is fundamentally limited by the system noise floor. When the interferometric intensity approaches the noise power density $N$, further reduction of the output cannot be experimentally resolved. The corresponding critical phase is therefore given by
    \begin{align}
        \phi_{0} = \frac{1}{2} \left[ \phi_{\rm{LO1}} + \phi_{\rm{LO2}} - \arccos{\left( 10^{N/10} -1\right)} \right].
        \label{eq9}
    \end{align}
    For convenience, the minimum detectable interferometric intensity variation is normalized to unity, i.e., $\delta I_{\rm IF}=1$. The corresponding minimum resolvable phase is therefore
    \begin{align}
        \delta \phi_{\rm{min}} = \frac{\delta I_{\rm{IF}}}{k_{\rm{IF0}}} = \frac{\ln 10}{20} \cot \left( \frac{\phi_{\rm{LO1}} + \phi_{\rm{LO2}} - 2\phi_{\rm{0}}}{2} \right),
        \label{eq11}
    \end{align}
    where $k_{\rm IF0}$ is the value of the transfer slope at the critical phase $\phi_0$. Equation~\ref{eq11} therefore establishes the theoretical phase-resolution limit of the proposed interferometer, showing that the ultimate resolution is jointly determined by the transfer slope and the system noise floor.

    The same transfer function can be exploited to enhance weak-field sensitivity by introducing an auxiliary microwave field that biases the interferometer at the critical interference point. The auxiliary and signal fields coherently combine before interacting with the atoms, yielding an equivalent field with resultant amplitude $E_t$ and phase perturbation $\delta \phi$,
    \begin{equation}
        \begin{aligned}
        E_{\rm{tot}} &= E_{\rm{Aux}} \cos(\omega t + \phi_{\rm{Aux}}) + E_{\rm{sig}} \cos(\omega t + \phi_{\rm{sig}}) \\
             &= E_{t} \cos(\omega t + \phi_{\rm{Aux}} + \delta \phi),
        \end{aligned}
        \label{eq12}
    \end{equation}
    from which the equivalent phase perturbation is obtained as
    \begin{align}
        \delta \phi = \arctan{\left[ \frac{\sin(\phi_{\rm{Aux}}) + x \sin(\phi_{\rm{sig}})}
                      { \cos(\phi_{\rm{Aux}}) + x \cos(\phi_{\rm{sig}})} \right]} - \phi_{\rm{Aux}}.
        \label{eq13}
    \end{align}
    In the weak-signal limit $x \ll 1$, Eq.~\ref{eq13} reduces to
    \begin{align}
        \delta\phi \approx x \sin(\phi_{\rm sig}-\phi_{\rm Aux}),
    \end{align}
    which reveals that the signal field induces a phase modulation with a sinusoidal dependence on the relative phase between the auxiliary and signal fields. The maximum attainable phase perturbation, obtained by optimizing over the signal phase, determines the largest interferometric modulation that can be produced by a signal of a given amplitude
    \begin{align}
        \delta \phi_{\rm{max}} = \arctan{\left[ \frac{\tan(\phi_{\rm{Aux}}) + x}
                      { 1 - x \tan(\phi_{\rm{Aux}})} \right]} - \phi_{\rm{Aux}}.
        \label{eq14}
    \end{align}
    Since the modulation depth is proportional to the maximum phase perturbation and inversely proportional to the residual interferometric background, the oscillation factor can be written as
    \begin{align}
        \kappa_{\rm{IF}} = 10 \log{\left( \frac{\delta \phi_{\rm{max}}}{I_{\rm{IF}}|_{\phi_{\rm{Aux}}}} \right)},
        \label{eq15}
    \end{align}
    which characterizes the modulation depth of the interferometric response, where $I_{\rm{IF}}|_{\phi_{\rm{Aux}}}$ is the normalized intensity of the interferometric component in the absence of the signal field, determined by the auxiliary field phase, with its lower bound constrained by the noise power density $N$. This quantity provides a direct theoretical metric for evaluating the attainable weak-field sensitivity and is used throughout the experimental analysis. Together, Eqs.~\ref{eq7}-\ref{eq15} establish the theoretical framework underlying the proposed dual-LO interferometer by quantitatively relating the interferometric transfer function, the attainable phase resolution, and the auxiliary-field-assisted sensitivity enhancement, thereby providing the theoretical basis for the experimental results presented in the main text.

    \textbf{Interferometric Polarization Response}
    The polarization response of the interferometer originates from the vector nature of the microwave-atom interaction. For a Rydberg transition with a defined quantization axis set by the optical-field geometry, the effective microwave coupling strength is determined by the projection of the microwave electric field onto the transition dipole moment.

    In our configuration, Horn1 defines a reference polarization axis, while the signal field radiated from Horn2 is rotated by an angle $\theta_{\rm sig}$ with respect to this axis, as shown in Figs.~\ref{fig5} (a). As a result, the effective field amplitude experienced by the atoms scales as $\cos(\theta_{\rm sig})$. Therefore, the polarization-dependent interferometric intensity can be expressed as
    \begin{align}
        I_{\rm Itf} = 10 \log \left \{ \left[1+\cos(\phi_{\rm Itf}) \right] \cdot \cos^2(\theta_{\rm sig}) \right \}.
        \label{eq16}
    \end{align}

    This expression indicates that the polarization dependence enters the interferometric response through a multiplicative amplitude modulation term proportional to $\cos^2(\theta_{\rm sig})$, while the phase-dependent interference term remains unchanged. Consequently, the interferometric response preserves the same functional structure as the phase transfer characteristic in Eq.~\ref{eq7}, establishing a direct correspondence between polarization and phase sensing within the same detection framework. As a result, polarization metrology inherits the same critical-point enhancement mechanism as phase sensing, as demonstrated experimentally in Figs.~\ref{fig5} (d)-(f).

    Overall, the polarization-to-intensity transduction extends the interferometric framework to an additional physical degree of freedom, enabling unified sensing of phase, amplitude, distance, and polarization within a single interferometric readout architecture.


\section*{Data Availability}
    All experimental data used in this study are available from the corresponding author upon request.
    
\section*{CODE AVAILABILITY}
    The custom codes used to produce the results presented in this paper are available from the corresponding authors upon request.

\section*{Author contributions statement}
    B.-B.W. conceived the idea. J.R.C conducted the physical experiments. The research was supervised by B.-B.W. All authors contributed to discussions regarding the results and the analysis contained in the manuscript.

\section*{Competing interests}
    The authors declare no competing interests.

\section*{References}
\bibliography{ref}

\end{document}